\documentclass[twocolumn,preprintnumbers,floatfix,pra,superscriptaddress]{revtex4-1}
\usepackage{amsmath}
\usepackage{amssymb}
\usepackage{graphicx}
\usepackage{dcolumn} 
\usepackage{ragged2e} 
\usepackage{bm}
\usepackage{epsfig}
\usepackage{hyperref}
\hypersetup{
    colorlinks=true,        
    linkcolor=blue,          
    citecolor=blue,        
    filecolor=magenta,      
    urlcolor=blue           
}
\usepackage{braket}
\usepackage{xcolor}

\begin{document} 
\title{Theoretical investigation of excitonic magnetism in LaSrCoO$_{4}$}

\author{J. Fern\'{a}ndez Afonso}
\affiliation{Institute of Solid State Physics,
TU Wien, Wiedner Hauptstr. 8, 1020 Vienna, Austria}
\author{A. Sotnikov}
\affiliation{Institute of Solid State Physics,
TU Wien, Wiedner Hauptstr. 8, 1020 Vienna, Austria}
\author{J. Kune\v{s}}
\email{kunes@ifp.tuwien.ac.at}
\affiliation{Institute of Solid State Physics,
TU Wien, Wiedner Hauptstr. 8, 1020 Vienna, Austria}
\affiliation{Institute of Physics, Czech Academy of Sciences, Na Slovance 2, 182 21 Praha 8, Czechia}

\date{\today}

\begin{abstract}
We use the LDA+U approach to search for possible ordered ground states of LaSrCoO$_4$. We find a staggered arrangement of magnetic multipoles to be stable over a broad range of Co $3d$ interaction parameters. This ordered state can be described as a spin-denity-wave-type condensate of $d_{xy} \otimes d_{x^2-y^2}$ excitons carrying spin $S=1$. 
Further, we construct an effective strong-coupling model, calculate the exciton dispersion and investigate closing of the exciton gap, which marks the exciton 
condensation instability. Comparing the layered LaSrCoO$_4$ with its pseudo cubic analog LaCoO$_3$, we find that for the same interaction parameters the 
excitonic gap is smaller (possibly vanishing) in the layered cobaltite.
\end{abstract}

\maketitle


\section{Introduction}\label{intro}
Perovskite cobaltites from the La$_{1-x}$Sr$_x$CoO$_3$ family has attracted much attention due their peculiar magnetic and transport
properties. In particular, the physics of the $x=0$ member with Co$^{3+}$ formal valence remains a subject of debate.
Recently, we have proposed that mobility of intermediate spin (IS) excitations plays an important role in the physics of
LaCoO$_3$~\cite{sotnikov16} and that the material is close to an excitonic condensation (EC) instability~\cite{mott61,keldysh65,halperin68}.
The EC scenario is also being discussed as the origin of the experimentally observed phase transition~\cite{tsubouchi04}, in 
the materials from Pr$_{0.5}$Ca$_{0.5}$CoO$_3$ family~\cite{kunes14b,nasu16,tatsuno16,yamaguchi17,sotnikov17}.

The layered cobaltites La$_{2-x}$Sr$_x$CoO$_4$ are much less studied although they may exhibit the similar physics.
The parent compound La$_2$CoO$_4$, formal valence Co$^{2+}$, is an antiferromagnetic insulator with a N\'eel 
temperature of 275~K~\cite{babkevich10}. The hole doping suppresses the N\'eel temperature and an incommensurate magnetic
order appears for $x>1/3$. The doping range $1/3<x<1/2$ is particularly interesting due to an hourglass-shaped spectrum
of magnetic excitations~\cite{drees14} invoking similarities to curate superconductors. The stoichiometric compound
LaSrCoO$_4$ with Co$^{3+}$ formal valence is relatively less studied due difficulties in sample preparation~\footnote{O. Kaman, private communication}.
The optical~\cite{moritomo95} and transport measurements~\cite{moritomo97,chichev06} for LaSrCoO$_4$ show a
good insulator with a charge gap of about 1~eV. The inverse magnetic susceptibility has a concave shape with
pseudo linear $T$-dependence below 150~K corresponding to $\mu_{\text{eff}}=2.3-2.6\,\mu_B$ and almost vanishing
Wiess temperature ranging from 27~K~\cite{guo16,chichev06} to 30~K~\cite{moritomo97}. The authors of 
Ref.~\cite{guo16} reported magnetic anomaly at 7~K, which they interpreted as formation of spin glass. The specific
heat data indicate only a modest entropy release of $0.06R$ at the supposed transition.

Theoretical studies of LaSrCoO$_4$ are even more limited. Wang {\it et al.}~\cite{wang00} applied the unrestricted Hartree--Fock approach
to a tight-binding model with Hubbard interaction in order to study various spin-state patterns and identified the high-spin--low-spin (HS--LS)
order to be the most likely ground state.

In this work, we use the LDA+U density-functional method to investigate possible ordered ground states of LaSrCoO$_4$, in 
particular the excitonic condensate. Furthermore, we employ the strong-coupling expansion to derive a low-energy
effective model in the Hilbert space spanned by the LS, IS, and HS states and compare it to an analogous model
of LaCoO$_3$. Similarly to our results for LaCoO$_3$, we find a stable excitonic ground state for realistic on-site
interaction parameters. Comparison to LaCoO$_3$ shows that for the same interaction parameters we obtain
a smaller or vanishing excitonic gap in LaSrCoO$_4$, which implies that LaSrCoO$_4$ is closer to the 
excitonic instability or the excitonic condensate is actually realized.

The paper is organized as follows: In Sec.~\ref{comp} we introduce the excitonic order parameter and explain the computational methods used in this work. In Sec.~\ref{res_lda} we present DFT calculations and investigate the stability of the excitonic condensate with and without spin-orbit coupling for different interaction parameters.
Sec.~\ref{res_sc} is devoted to the strong-coupling analysis of dispersion of bosonic excitations in the 
normal state, the softening of which marks the onset of condensation.

\section{Computational method}\label{comp}
\subsection{LDA+U}

Electronic structure calculations in this paper are performed in the framework of the density functional theory (DFT) by using local density approximation (LDA)~\cite{ksdft,hklda}. The effect of Coulomb interaction within the $3d$ shell of Co is described by means of LDA+U scheme with the so-called fully localized limit  double-counting correction~\cite{Fully_loc_limmit}. The spin polarization enters the orbital-dependent potential only while leaving the LDA exchange-correlation potential unpolarized. The calculations are carried out with the WIEN2k~\cite{wien2k} package.

\begin{figure}
\includegraphics[width=\columnwidth]{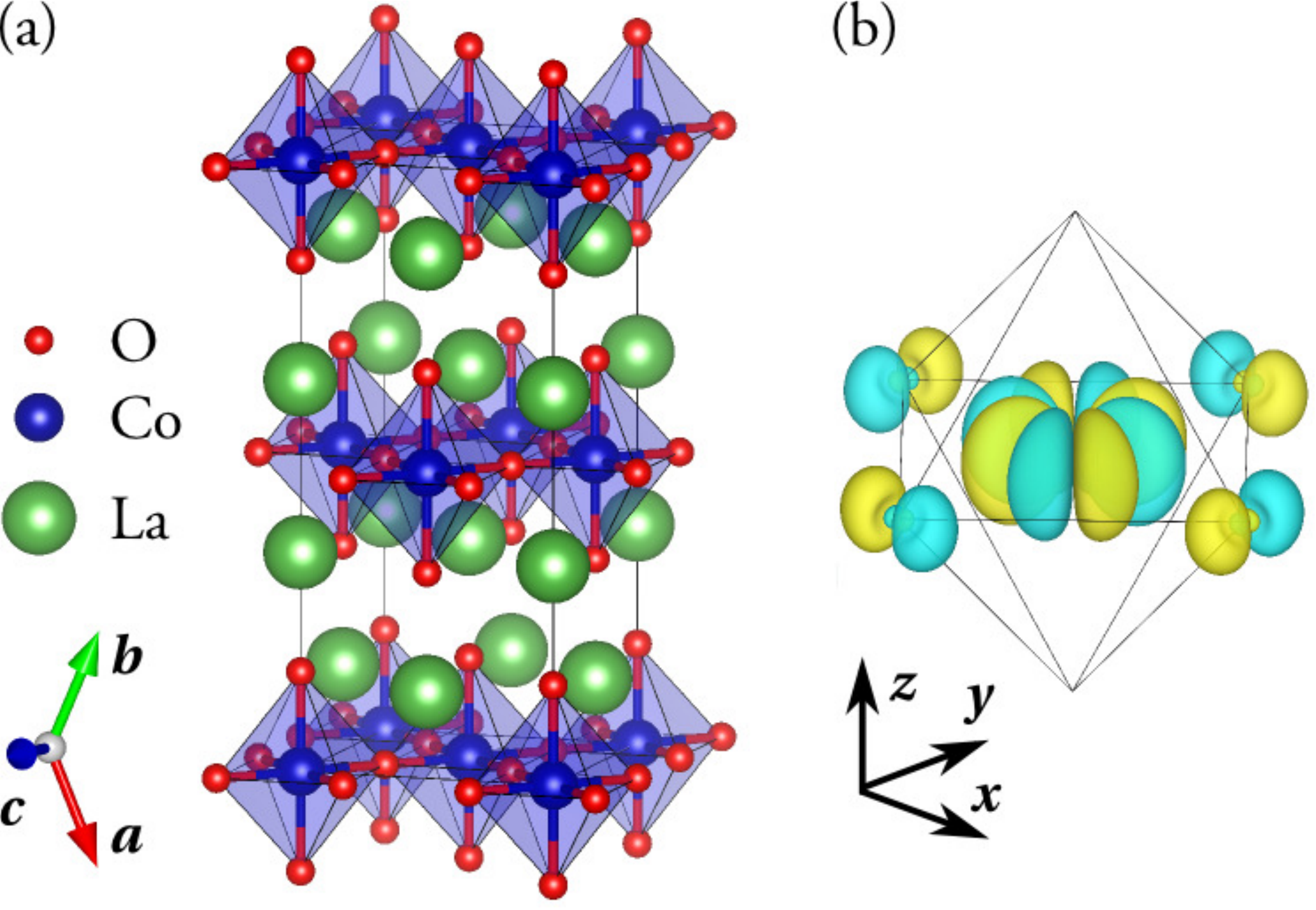}
\caption{(a) Conventional unit cell of La$_2$CoO$_{4}$. On the left side the coordinate basis $(\boldsymbol{a},\boldsymbol{b},\boldsymbol{c})$ for the  corresponding primitive unit cell with 2~f.u. is shown. 
(b) Isosurface of the excitonic spin-density distribution obtained around the cobalt atom. Blue and yellow colors represent the positive and negative spin projections, respectively. The spin density around oxygen atoms indicates that the exciton propagates only in the CoO$_6$ planes.}
\label{structure}
\end{figure}

The La$_{2-x}$Sr$_x$CoO$_{4}$ structure consists of single layers of CoO$_6$ corner-sharing octahedra separated by a random distribution of La and Sr ions (see Fig.~\ref{structure}(a)). 
To simulate the LaSrCoO$_{4}$ compound, we employ the virtual crystal approximation (VCA)~\cite{nordheim31, muto38} that sets the atomic number of La to 56.5.
Anticipating the possibility of the staggered order~\cite{afonso17}, we choose a unit cell allowing for the $\sqrt{2} \times \sqrt{2}$ in-plane order. The unit cell vectors shown in Fig.~\ref{structure}(a) correspond to the structural parameters $a = 6.8019$~\AA, $b = 6.8019$~\AA, $c = 5.3796$~\AA, $\alpha = \pi/2$, $\beta = \pi/2$, and $\gamma = 2.3284$. The Brillouin zone is sampled with the uniform $4 \times 4 \times 4$ $k$ mesh. The muffin-tin radii (in bohrs) are the following: 2.50 for La, 1.91 for Co, and 1.65 for O. The plane wave cutoff is set to $R_{mt}K_{max} = 6$.

In search for possible ordered solutions we start from several different initial states with broken symmetry: the staggered excitonic and spin-state (LS--HS) orders. The calculations are performed with and without spin-orbit coupling (SOC). The anticipated EC order with the $d_{x^2 - y^2} \otimes d_{xy}$ orbital symmetry is described by a three-dimensional complex vector~$\boldsymbol{\phi}$ with components
\begin{equation}\label{ordparam}
\phi_{\beta} = \langle \hat{O}'_\beta \rangle + i \langle \hat{O}''_\beta \rangle = \phi'_{\beta} + i \phi''_{\beta},
\end{equation}
expressed in terms of the hermitian operators 
\begin{eqnarray}
  \hat{O}'_{\beta}=\frac{i}{4}\sum\limits_{\sigma,\sigma'=\uparrow,\downarrow}(\tau_{\beta})_{\sigma\sigma'}(\hat{c}^{\dagger}_{2\sigma}\hat{c}^{\phantom\dagger}_{-2\sigma'} - \hat{c}^{\dagger}_{-2\sigma}\hat{c}^{\phantom\dagger}_{2\sigma'}),
  \nonumber
  \\
  \hat{O}''_{\beta}=\frac{1}{4}\sum\limits_{\sigma,\sigma'=\uparrow,\downarrow}(\tau_{\beta})_{\sigma\sigma'}(\hat{c}^{\dagger}_{-2\sigma}\hat{c}^{\phantom\dagger}_{-2\sigma'} - \hat{c}^{\dagger}_{2\sigma}\hat{c}^{\phantom\dagger}_{2\sigma'}),
  \label{O_operator}
\end{eqnarray}
where $\hat{c}^{\dagger}_{m\sigma}$ ($\hat{c}^{\phantom\dagger}_{m\sigma}$) are the creation (annihilation) operators for $3d$ electrons on the same Co atom (the site indices are not shown for sake of simplicity). The orbital index $m$ refers to the spherical harmonic $Y_{2,m}$ basis with the $z$ quantization axis (see Fig.~\ref{structure}(b)). The Pauli matrices  $\tau_{\beta}$ ($\beta=x,y,z$) capture the spin-triplet character of the EC order. A nonzero $\boldsymbol{\phi}'$ gives rise to a spin-density distribution on the Co site with vanishing spin moment per atom that is shown in Fig.~\ref{structure}(b). For a nonzero $\boldsymbol{\phi}''$, the spin density is identically zero, while the spin-rotational symmetry is broken due to presence of local spin currents.

\subsection{Strong coupling expansion}\label{SCexp}

The starting point of our strong-coupling analysis is the Hubbard Hamiltonian for the $d$-shells of Co
\begin{equation}\label{ham_FH}
 \hat{\cal H} =
 \sum_{i}\hat{\cal H}^{(i)}_{\text{at}}
 +\sum_{{\bf r}}\hat{\cal H}^{({\bf r})}_{t},
\end{equation}
where
\begin{eqnarray}
&&\hat{\cal H}^{(i)}_{\text{at}} =
\sum_{\alpha\beta}h_{\alpha\beta}^{ii}\hat{c}^\dag_{i\alpha}
\hat{c}_{i\beta}\nonumber
\\
&&\qquad\qquad
 + \sum_{\alpha\beta\gamma\delta}U_{\alpha\beta\gamma\delta}
\hat{c}^\dag_{i\alpha}\hat{c}^\dag_{i\beta}\hat{c}_{i\gamma}\hat{c}_{i\delta},
\label{ham_int}
\\
&&\hat{\cal H}^{({\bf r})}_{t}\equiv\hat{\cal H}^{(ij)}_{t}=\sum_{\alpha\beta}
h_{\alpha\beta}^{ij}\hat{c}^\dag_{i\alpha}
\hat{c}_{j\beta},~i\neq j,
\label{ham_hop}
\end{eqnarray}
$i$ and $j$ are now the lattice-site indices (below, we account for local, $i=j$, 
and nearest-neighbor contributions only) that correspond to a particular bond index ${\bf r}$ for $i\neq j$, 
and $\alpha,\beta,\gamma,\delta$ are the internal state (orbital and spin) indices. 
The local and nearest-neighbor hopping matrices $h_{\alpha\beta}^{ij}$ are provided
by projection of the LDA band structure to Wannier orbitals. 
To this end, we use wien2wannier~\cite{wien2wannier} and wannier90~\cite{wannier90} software.
We use the Slater parameterization of the intra-atomic electron-electron interaction $U_{\alpha\beta\gamma\delta}$ of two tuneable parameters, 
average interaction~$\widetilde{U}$ and Hund's coupling~$\widetilde{J}$, while fixing the ratio of Slater integrals $F_4/F_2= 0.625$.
While the parametrization of the local Coulomb interaction on Co is the same as the one employed in the LDA+U method, the values of the effective 
interaction parameters should be different since they refer to a different model. 

We diagonalize the local Hamiltonian $\hat{\cal H}^{(i)}_{\text{at}}$ to obtain atomic eigenenergies~$E^{(q)}_{\gamma}$ and 
eigenstates~$\ket{\Psi^{(q)}_{\gamma}}$, where $q$ is the number of electrons in the $d$-shell and $\gamma$ is the state index.
Next, we use the set of the lowest 25 states of the $d^6$ configuration containing LS, IS, and HS as an active space 
and treat the non-local terms $\hat{\cal H}^{({\bf r})}_t$ as a perturbation.
Performing the Schrieffer--Wolff transformation~\cite{sw66} to the second order, we arrive at the following 
bosonic Hamiltonian:
\begin{eqnarray}\label{ham_BH}
 \hat{\cal H}_{\text{eff}} &=&
 \sum_{ij,\,\alpha\beta}(\varepsilon_{1\alpha\beta}^{ij}
 \hat{d}^\dag_{i\alpha}\hat{d}_{j\beta}
 \hat{s}_{i}\hat{s}^\dag_{j}
 + \varepsilon_{2\alpha\beta}^{ij}
 \hat{d}^\dag_{i\alpha}\hat{d}^\dag_{j\beta}\hat{s}_{i}\hat{s}_{j})
 \nonumber
 \\
 && + {\text{H.c.}} + \hat{\cal H}_{\text{int}}.
\end{eqnarray}
Here, we consider the LS state as the bosonic vacuum, $\ket{\emptyset}_i=\hat{s}_i^\dagger\ket{0}$, and other states
from the low-energy manifold of the $d^6$ configuration as different bosonic flavors $\alpha$ characterized by the corresponding
creation (annihilation) operators $\hat{d}^{\dag}_{i\alpha}$ ($\hat{d}_{i\alpha}$) on the lattice site $i$.

In Eq.~\eqref{ham_BH} we distinguish three types of terms. The first one with the amplitude $\varepsilon_{1}$
corresponds to the {\it renormalized} on-site energies of bosons (for $i=j$),
\begin{eqnarray}\label{eloc}
 \varepsilon_{1\alpha\beta}^{ii}
  =E_\alpha^{(6)}\delta_{\alpha\beta}+\sum_{{\bf r},v w,\nu=\pm1}
  {{\cal M}^{({\bf r})(6+\nu)}_{\alpha\beta, w v,\alpha\beta}}/{{\cal E}^{(6+\nu)}_{\alpha\beta, w v}}\,,
\end{eqnarray}
and their hopping amplitudes on the LS background (for $i\neq j$),
\begin{eqnarray}\label{ehop}
 \varepsilon_{1\alpha\beta}^{ij}
  =\frac{1}{2}\sum_{v w,\nu=\pm1}
  \left(\frac{1}{{\cal E}^{(6+\nu)}_{\emptyset\beta, w v}}+\frac{1}{{\cal E}^{(6+\nu)}_{\alpha\emptyset, w v}}\right)
  {{\cal M}^{(ij)(6+\nu)}_{\emptyset\beta, w v,\alpha\emptyset}}\,.
\end{eqnarray}
The second term with the amplitude $\varepsilon_{2}$ corresponds to the non-local pair-creation processes, 
\begin{eqnarray}\label{epair}
 \varepsilon_{2\alpha\beta}^{ij}
  =\frac{1}{2}\sum_{v w,\nu=\pm1}
  \left(\frac{1}{{\cal E}^{(6+\nu)}_{\emptyset\emptyset, w v}}+\frac{1}{{\cal E}^{(6+\nu)}_{\alpha\beta, w v}}\right)
  {{\cal M}^{(ij)(6+\nu)}_{\emptyset\emptyset, w v,\alpha\beta}}\,,
\end{eqnarray}
where
\begin{eqnarray}\label{Mr}
 {\cal M}^{({\bf r})(6\pm1)}_{\gamma\delta, w v,\gamma'\delta'}=
 \bra{\Psi^{(6)}_{\gamma'}\Psi^{(6)}_{\delta'}}
 \hat{\cal H}^{({\bf r})}_t
 \ket{\Psi^{(6\pm1)}_{v}\Psi^{(6\mp1)}_{w}}
 \nonumber
 \\
 \times
 \bra{\Psi^{(6\pm1)}_{v}\Psi^{(6\mp1)}_{w}}
 \hat{\cal H}^{({\bf r})}_t
 \ket{\Psi^{(6)}_{\gamma}\Psi^{(6)}_{\delta}},
\end{eqnarray}
and ${\cal E}^{(6\pm1)}_{\gamma\delta, w v}=E^{(6)}_{\gamma}+E^{(6)}_{\delta}-E^{(6\pm1)}_{v}-E^{(6\mp1)}_{w}$.

The last term $\hat{\cal H}_{\text{int}}$ in Eq.~\eqref{ham_BH} characterizes all other processes with nonzero amplitudes 
that appear due to Schrieffer--Wolff transformation. 
Note that the main contributions to this term originate from (density-density and magnetic/orbital exchange) interactions between different bosonic flavors. 
Including the effect of  $\hat{\cal H}_{\text{int}}$ on the exciton dispersion/spectra is a non-trivial task and goes beyond the scope of the current paper.

Here, we focus on the excitonic instability of the normal ground state that has dominantly the LS character.
The small $\epsilon_2$ allows us to neglect the density of $d$-bosons in the ground state as well as
the in the lowest excited states. The elementary excitations are then approximately described by the bilinear part of 
the Hamiltonian~\eqref{ham_BH}, neglecting higher-order terms (with three and four operators $\hat{d}_\alpha$) contributing to $\hat{\cal H}_{\text{int}}$.
Using the same argument we drop the hard-core constraint on $d$ bosons and proceed with the linearized spin-wave approach~\cite{svb01}, 
which provides access to momentum dependencies of bosonic (IS and HS) excitations in the lattice. 
The excitation spectrum obtained in this way corresponds to one-boson excitations of the normal ground state, 
thus cannot account for effects originating from thermal occupation of IS and HS states.

\section{Results}\label{res}
\subsection{LDA+U calculations}\label{res_lda}

LDA+U calculations without SOC lead to a non-zero $\boldsymbol{\phi}'$ with staggered arrangement in the Co-O plane. We have verified that different spatial orientations of $\boldsymbol{\phi}'$ are numerically equivalent. The ground state can be described as a polar excitonic condensate~\cite{ho98}, or, more specifically, as a spin-density wave excitonic condensate in the classification of Halperin and Rice~\cite{halperin68}. The corresponding spin-density distribution around Co site, which gives vanishing atomic moment, is shown in Fig.~\ref{structure}(b).

\begin{figure}
\includegraphics[width=\columnwidth]{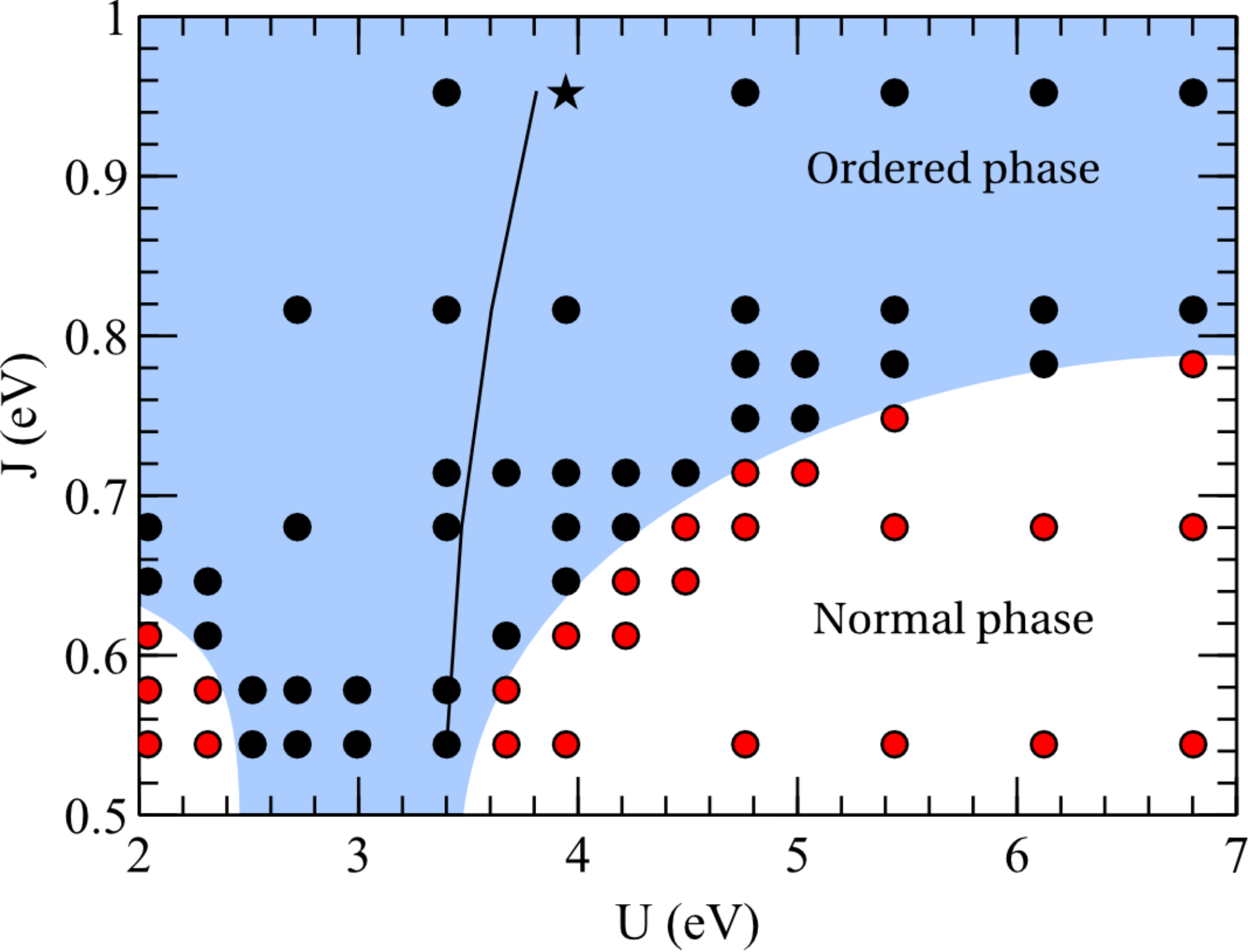}
\caption{The $U$-$J$ phase diagram indicating the local stability of the EC state obtained by the LDA+U approach. 
The symbols correspond to the actual calculations performed.
The black line indicates the metal-insulator transition in constrained normal phase.}
\label{UJ_phase}
\end{figure}

\begin{figure}
\includegraphics[width=\columnwidth]{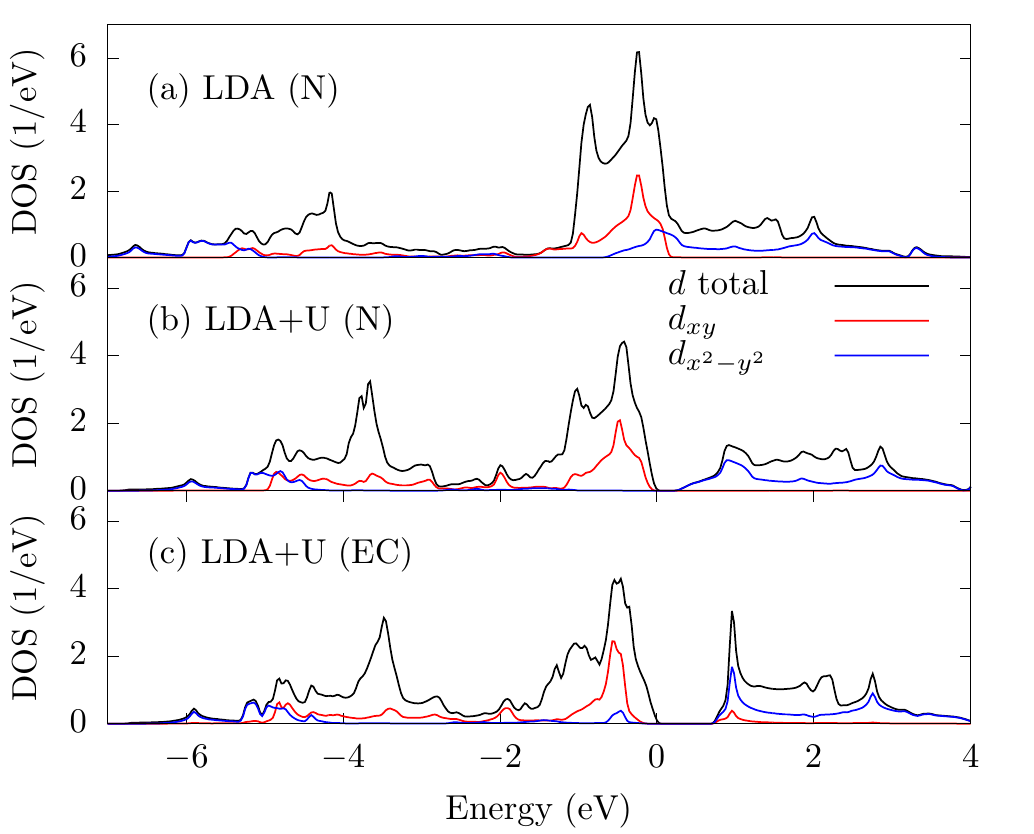}
\caption{The spectral density of Co $d$ states for the (a) LDA normal (N) solution, (b) LDA+U normal phase and (c) LDA+U EC phase obtained with  $U = 3.95$ eV and $J = 0.95$ eV. Notice the hybridization between $d_{xy}$ and $d_{x^2-y^2}$ states at low energies in the EC solution.}
\label{DOS}
\end{figure}

The stability of the EC long-range order as a function of $U$ and $J$ is summarized in  Fig.~\ref{UJ_phase}. 
The overall shape of the phase diagram resembles the results for cubic LaCoO$_3$~\cite{afonso17}. We find ordered solutions in both the weak coupling (metallic) and the strong coupling regimes with a wedge of the normal phase at intermediate $U$ and small $J$. The modification of spectral density due to the excitonic condensation is shown in Fig.~\ref{DOS}.

Introducing SOC breaks the spin-rotational symmetry. The non-zero value of $\langle xy , \sigma | \hat{l}_{z}\hat{s}_{z} |x^2-y^2 , \sigma \rangle$ can be viewed as a source field for $\phi_{z}''$~\cite{afonso17}. Upon inclusion of SOC, the EC state splits into two distinct stable solutions 
\begin{eqnarray}
  &&\boldsymbol{\phi}^{\parallel}_{j} 
  = (-1)^{j}(\phantom{\lambda'_{\bot}}\hspace{-0.6em}0~\phantom{\lambda'_{\bot}}\hspace{-0.6em}0~\lambda'_{\parallel})
  + i (0~0~\lambda''_{\parallel}),
  \nonumber
  \\
  &&\boldsymbol{\phi}^{\bot}_{j} 
  = (-1)^{j}(\lambda'_{\bot}~\lambda'_{\bot}~0)
  + i (0~0~\lambda''_{\bot}),
  \label{order_SOC}
\end{eqnarray}
where $(-1)^j$ indicates the staggered in-plane arrangement. The $\boldsymbol{\phi}^{\parallel}$ and $\boldsymbol{\phi}^{\bot}$ states have different symmetries, reflected, for example, in the appearance of small but finite staggered moment ${\bf m}_{j} \propto \boldsymbol{\phi}'_{j} \times \boldsymbol{\phi}''_{j}$ of $0.05$ $\mu_B$ for $\boldsymbol{\phi}^{\bot}\neq0$. No ordered moment arises in the $\boldsymbol{\phi}^{\parallel}$ solutions. The total energies shown in Table~\ref{order_table} for $U = 3.95$~eV and $J = 0.95$~eV (the star-shaped point in Fig.~\ref{UJ_phase}) favor the $\boldsymbol{\phi}^{\parallel}$ order. 
In this parameter regime we have also searched for other possible two sublattices solutions such as antiferromagnetic or spin-state order (LS-HS)~\cite{wang00}. All our calculations converged to the $\boldsymbol{\phi}^{\parallel}$ solutions. The calculations were performed without lattice relaxation. 
\begin{table}
\begin{tabular}{lcccc}
\hline
\hline\\
Solution & Approach & $| \boldsymbol{\phi} ' |$ & $| \boldsymbol{\phi} '' |$ & $\Delta E$~(meV/f.u.)\\[0.5ex]
\hline\\
EC & LDA+U &  0.289 & 0.000 &  -14.25 \\[0.5ex]
EC$^{\bot}$ &~~LDA+U+SOC~~& 0.274 & 0.098 &  -14.83\\[0.5ex]
EC$^{\parallel}$ &~~LDA+U+SOC~~& 0.299 & 0.096 &  -17.58\\[0.5ex]
\hline
\hline
\end{tabular}
\caption{Comparison of the order parameters $| \boldsymbol{\phi} ' |$ and $| \boldsymbol{\phi} ''|$ and the energy difference $\Delta E$ (compared to the normal state) of the ordered solutions at $U=3.95$~eV and $J=0.95$~eV.}
\label{order_table}
\end{table}

\subsection{Strong coupling analysis}\label{res_sc}

\begin{figure}
\includegraphics[width=\columnwidth]{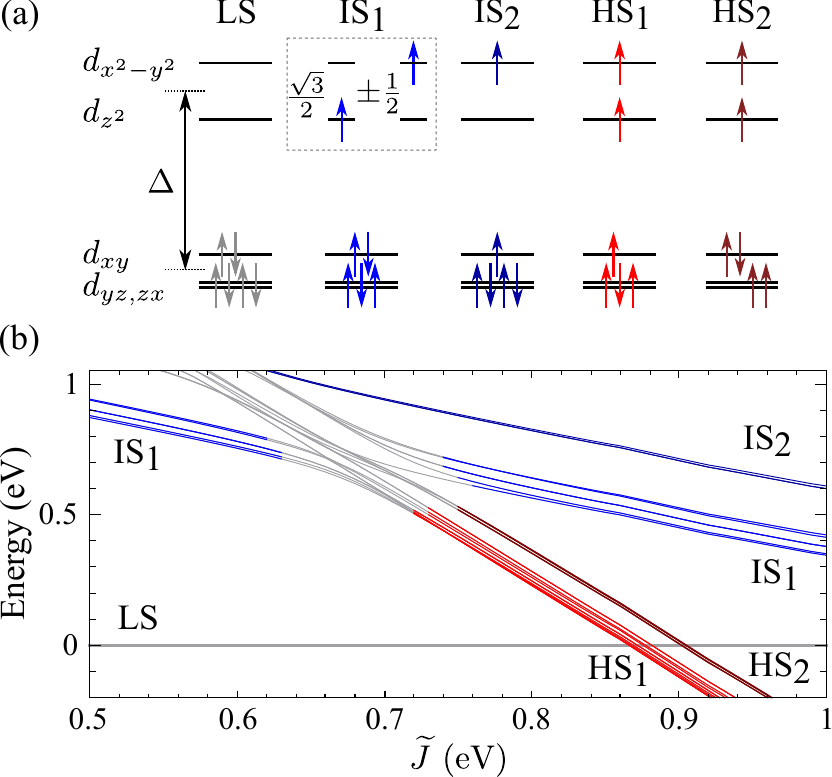}
\caption{Spin configurations of the lowest atomic multiplets of La$_2$CoO$_4$ (a) and the
corresponding atomic multiplet energies for SOC $\zeta=56$\,meV (b).}
\label{TS}
\end{figure}

In Fig.~\ref{TS} we show the set of eigenenergies $E^{(6)}_\gamma$ of the Hamiltonian~\eqref{ham_int} corresponding to the lowest atomic multiplets of LaSrCoO$_4$ as a function of the Hund's coupling~$\widetilde{J}$.
The three intermediate-spin $d_{xy}\otimes d_{x^2-y^2}$ states (IS$_2$) have the highest atomic energies
among 25 lowest states in the region of realistic values of Hund's coupling~$\widetilde{J}$ due 
to cubic crystal field ($\Delta=1.594$\,eV) and additional tetragonal spittings of the $t_{2g}$ and $e_{g}$ states (0.103\,eV and  0.384\,eV, respectively). 
This seems to prevent the IS$_2$ states from condensation. However, as we show below,
the renormalization of on-site energies by virtual hopping processes reverses the order of atomic multiplets, i.e.,
when placed on the lattice the IS$_2$ excitations have lower energies than the IS$_1$ ones.

Next, we compute the amplitudes \eqref{eloc}--\eqref{epair} and
diagonalize the bilinear part of the effective Hamiltonian~\eqref{ham_BH}.
\begin{figure}
\includegraphics[width=\columnwidth]{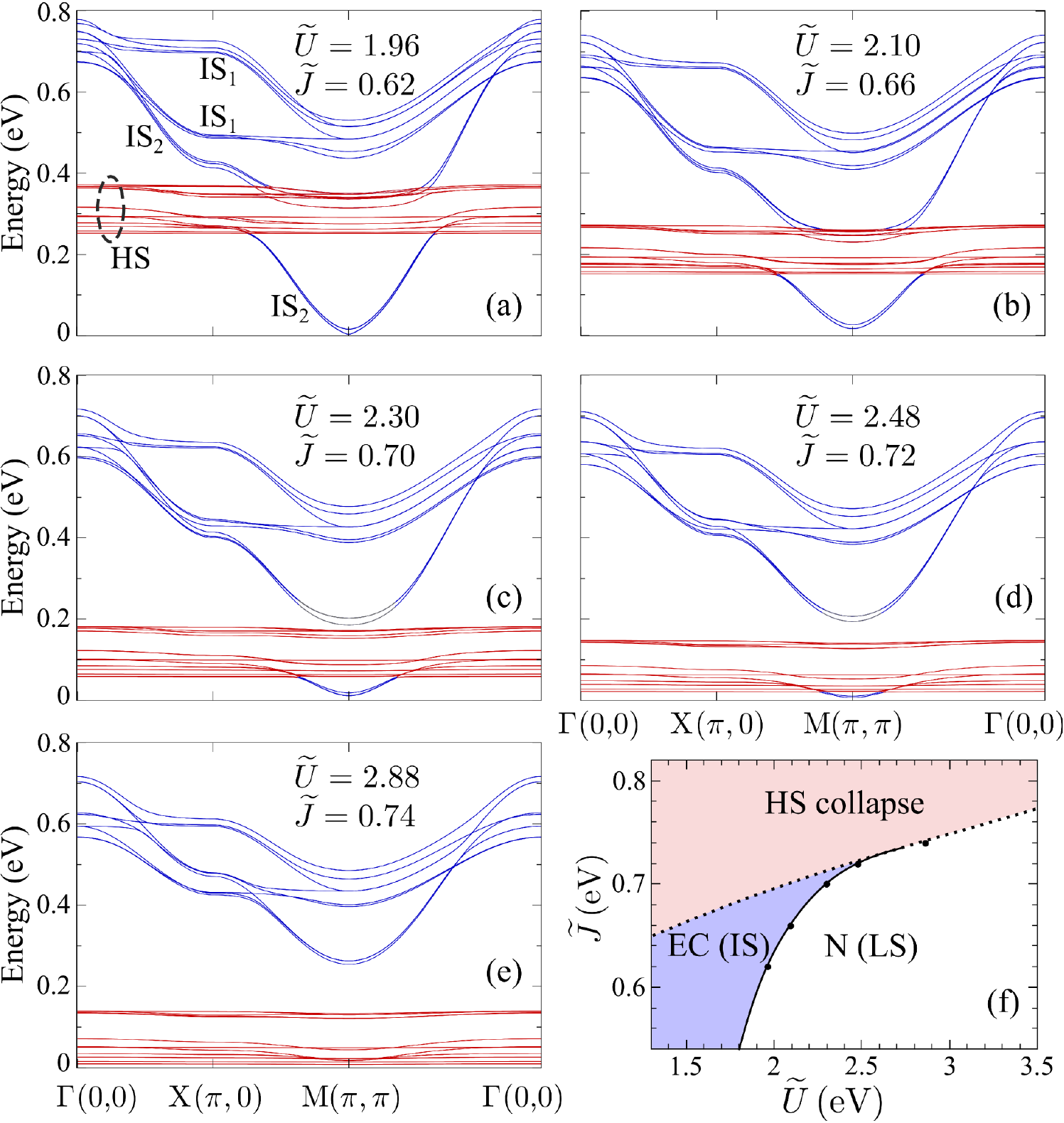}
  \caption{(a)--(e) Dispersions of the IS and HS excitations at different $\widetilde{U}$ and $\widetilde{J}$ (given in eV) and $\zeta=56$\,meV. (f) The corresponding 
  phase diagram obtained following the disappearance of the exciton gap.
  }\label{disp}
\end{figure}
In Fig.~\ref{disp}(a)--(e) we show the dispersions of elementary 
(IS- and HS-like) excitations of~\eqref{ham_BH} obtained for different 
values of $\widetilde{U}$ and $\widetilde{J}$. While the IS excitations can move on the LS background, the HS excitations cannot, and
the deviation from completely flat HS bands is due to HS-IS mixing through spin-orbit coupling.
Note that the spin-orbital character of the excitations remains approximately fixed along the individual branches
due to conservation of orbital and spin flavors in the nearest-neighbor kinetic exchange processes. Small mixing and 
thus momentum dependence arises due to spin-orbit coupling.

The interaction parameters were chosen so that the solution is at the verge of excitonic instability, i.e., 
normal solutions with a tiny gap, vanishing of which determines the phase boundary in Fig.~\ref{disp}(f).
Variation of $\widetilde{J}$ enters predominantly through the changes of atomic multiplet energies in Fig.~\ref{TS}.
Variation of $\widetilde{U}$ affects both the hopping amplitudes (bandwidths) and the renormalization of
site energies (band shifts). The renormalizations of different states scale differently with $\widetilde{U}$ 
depending of the number and amplitudes of the virtual hopping process for a given state on a LS background.

\begin{figure}
\includegraphics[width=\columnwidth]{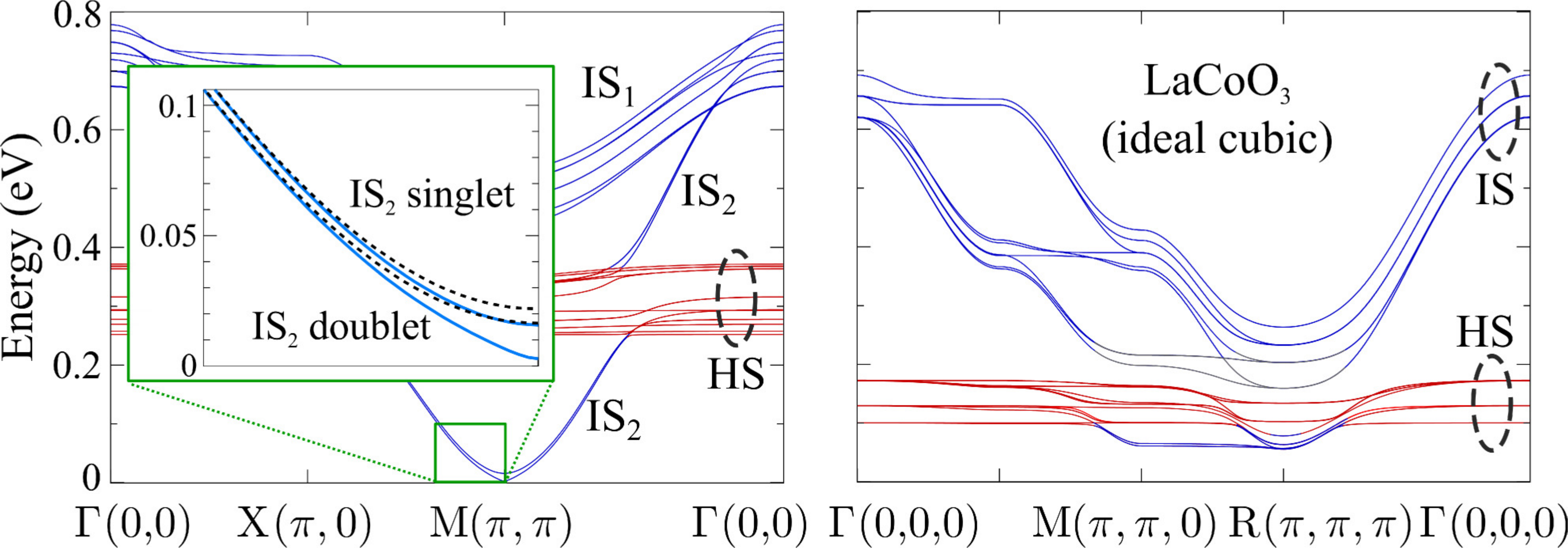}
  \caption{Dispersions of the IS and HS excitations for LaSrCoO$_4$ (left) and cubic LaCoO$_3$ (right)
  at $\widetilde{U}=1.96$~eV, $\widetilde{J}=0.62$~eV, and $\zeta=56$~meV.
  Inset: IS$_2$ dispersion with (solid lines) and without (dashed lines) pair-creation contributions. 
  }\label{lsco-laco}
\end{figure}
The influence of the pair-creation term  is shown in the inset of Fig.~\ref{lsco-laco}. 
Due to bosonic nature of excitations, this term effectively acts as an ``attraction'' between bands $\epsilon_\gamma({\bf k})$
and their mirror images $-\epsilon_\gamma({\bf k})$. 
Only when their separation is comparable or smaller than characteristic amplitudes determined by Eq.~\eqref{epair}, the pair creation/annihilation processes become important.
These amplitudes for the cases under study are typically of the order of few meV, thus the effect becomes noticeable close to the minimum of the IS$_2$ energy dispersion.

To compare with LaCoO$_3$, we repeat the analysis 
for the hypothetical cubic structure of Ref.~\cite{afonso17} and the same $\widetilde{U}$ and $\widetilde{J}$ values,
see Fig.~\ref{lsco-laco}. 
We find the excitation gap is larger in the cubic structure due to narrower IS band.
Moreover, the HS excitations in the layered structure are located at substantially higher energy than in its cubic counterpart.
This is because the out-of-plane nearest-neighbor processes, which contribute to renormalization of the HS local energies in the cubic
structure, are missing in the layered system. Similar processes play a minor role for the IS$_2$ excitations and thus the renormalizations
in the cubic and layered structures are comparable. 

More extensive numerical analysis confirms the general tendency that with a continuous decrease of $\widetilde{U}$ and $\widetilde{J}$ parameters,
the EC instability first appears in the layered compound and only then in LaCoO$_3$.
This agrees well with comparison of phase boundaries for critical $U$ and $J$ values in Fig.~\ref{UJ_phase}
with the values for LaCoO$_3$ published in Ref.~\cite{afonso17} that are both obtained within the LDA+U approach.

\section{Conclusions}

We have performed LDA+U calculations for LaSrCoO$_{4}$ that amount to a static mean-field treatment of the excitonic order in this compound. We find excitonic condensate to be a stable solution with a total energy lower than the one of normal state over a large part of the studied $U$-$J$ phase diagram. The generally used $U$ and $J$ values fall close to the boundary of the region of EC stability. The stable EC solutions are of the spin-density-wave type with a $d_{xy} \otimes d_{x^2-y^2}$ orbital symmetry and out-of-plane spin polarization. Comparison to LaCoO$_{3}$ by means of the linear spin-wave treatment of the effective strong-coupling model suggest that the layered cobaltite is closer to the excitonic instability or that the EC order is possibly realized. Investigations of behaviour under pressure and in high magnetic fields~\cite{ikeda} are highly desirable.  

\section{Acknowledgments}
The authors thank A. Hariki, O. Kaman and L. H. Tjeng for fruitful discussions.
This work has received funding from the European Research Council (ERC) under the European Union's Horizon 2020 research
and innovation programme (grant agreement No. 646807-EXMAG).
Access to computing and storage facilities provided by the Vienna Scientific Cluster (VSC) is greatly appreciated.


%

\end{document}